\documentstyle[12pt]{article}
\catcode`\@=11
\def\marginnote#1{}
\def\ifmath#1{\relax\ifmmode #1\else $#1$\fi}

\def\bold#1{\setbox0=\hbox{$#1$}%
     \kern-.025em\copy0\kern-\wd0
     \kern.05em\copy0\kern-\wd0
     \kern-.025em\raise.0433em\box0 }

\def\GENITEM#1;#2{\par\vskip6pt \hangafter=0 \hangindent=#1
   \Textindent{$ #2$ }\ignorespaces}

\newcount\hour
\newcount\minute
\newtoks\amorpm
\hour=\time\divide\hour by60
\minute=\time{\multiply\hour by60 \global\advance\minute by-
\hour}
\edef\standardtime{{\ifnum\hour<12 \global\amorpm={am}%
    \else\global\amorpm={pm}\advance\hour by-12 \fi
    \ifnum\hour=0 \hour=12 \fi
    \number\hour:\ifnum\minute<100\fi\number\minute\the\amorpm}}
\edef\militarytime{\number\hour:\ifnum\minute<100\fi\number\minute}
\def\draftlabel#1{{\@bsphack\if@filesw {\let\thepage\relax
  \xdef\@gtempa{\write\@auxout{\string
    \newlabel{#1}{{\@currentlabel}{\thepage}}}}}\@gtempa
    \if@nobreak \ifvmode\nobreak\fi\fi\fi\@esphack}
     \gdef\@eqnlabel{#1}}
\def\@eqnlabel{}
\def\@vacuum{}
\def\draftmarginnote#1{\marginpar{\raggedright\scriptsize\tt#1}}
\def\draft{\oddsidemargin -.5truein
        \def\@oddfoot{\sl preliminary draft \hfil
        \rm\thepage\hfil\sl\today\quad\militarytime}
        \let\@evenfoot\@oddfoot \overfullrule 3pt
        \let\label=\draftlabel
        \let\marginnote=\draftmarginnote

\def\@eqnnum{(\theequation)\rlap{\kern\marginparsep\tt\@eqnlabel}%
\global\let\@eqnlabel\@vacuum}  }
\def\preprint{\twocolumn\sloppy\flushbottom\parindent 1em
        \leftmargini 2em\leftmarginv .5em\leftmarginvi .5em
        \oddsidemargin -.5in    \evensidemargin -.5in
        \columnsep 15mm \footheight 0pt
        \textwidth 250mmin      \topmargin  -.4in
        \headheight 12pt \topskip .4in
        \textheight 175mm
        \footskip 0pt

\def\@oddhead{\thepage\hfil\addtocounter{page}{1}\thepage}
        \let\@evenhead\@oddhead \def\@oddfoot{} \def\@evenfoot{}
}
\def\titlepage{\@restonecolfalse\if@twocolumn\@restonecoltrue\onecolumn
     \else \newpage \fi \thispagestyle{empty}\c@page\z@
        \def\thefootnote{\fnsymbol{footnote}} }
\def\endtitlepage{\if@restonecol\twocolumn \else  \fi
        \def\thefootnote{\arabic{footnote}}
        \setcounter{footnote}{0}}  
\catcode`@=12
\relax

\def\be{\begin{equation}}
\def\ee{\end{equation}}
\def\bea{\begin{eqnarray}}
\def\eea{\end{eqnarray}}
\def\simlt{\stackrel{<}{{}_\sim}}

\def\NPB#1#2#3{{\it Nucl.~Phys.} {\bf{B#1}} (19#2) #3}
\def\PLB#1#2#3{{\it Phys.~Lett.} {\bf{B#1}} (19#2) #3}
\def\PRD#1#2#3{{\it Phys.~Rev.} {\bf{D#1}} (19#2) #3}

\def\PR#1#2#3{{\it Phys.~Rep.} {\bf#1} (19#2) #3}

\def\mst1{m_{\;\widetilde{t}_{1}}}

\def\mst2{m_{\;\widetilde{t}_{2}}}
\def\mst12{m_{\;\widetilde{t}_{1,2}}}

\def\msb1{m_{\;\widetilde{b}_{1}}}
\def\msb2{m_{\;\widetilde{b}_{2}}}
\def\msb12{m_{\;\widetilde{b}_{1,2}}}

\def\mtilde2{\widetilde{m}^{2}}

\relax

\begin{document}
\topmargin-2.5cm

%
\begin{titlepage}
\begin{flushright}
IEM-FT-175/98 \\
UAB-FT-449 \\
hep--ph/9806263 \\
\end{flushright}
\vskip 0.3in
\begin{center}{\Large\bf The Standard Model from extra 
dimensions~\footnote{Work supported in part by the CICYT of Spain
(contracts AEN95-0195 and AEN95-0882).} }
\vskip .5in
{\bf A. Pomarol$^{\dagger}$ and  M. Quir{\'o}s$^{\ddagger}$} \\
\vskip.35in
$\dagger$~{Institut de F{\'\i}sica d'Altes Energies,}
{Universitat Aut{\`o}noma de Barcelona,}\\
{E-08193 Bellaterra, Barcelona, Spain}\\  \vspace{.4cm}
$^\ddagger$~Instituto de Estructura de la Materia (CSIC), 
Serrano 123, 28006-Madrid, Spain
\end{center}
\vskip2.cm
\begin{center}
{\bf Abstract}
\end{center}
\begin{quote}
We present a simple $N=1$ five-dimensional model where the fifth dimension
is compactified on the orbifold $S^1/{\bf Z}_2$. Non-chiral matter lives 
in the bulk of the fifth dimension (five dimensions) while chiral matter 
lives on the fixed points of the orbifold (four-dimensional boundaries).
The massless sector constitutes the Minimal Supersymmetric Standard Model 
while the massive modes rearrange in $N=2$ supermultiplets. After 
supersymmetry breaking by the Scherk-Schwarz mechanism the zero modes can 
be reduced to the non-supersymmetric Standard Model. 
\end{quote}

\vskip2.cm

\begin{flushleft}
IEM-FT-175/98\\
June 1998 \\
\end{flushleft}
\end{titlepage}
\setcounter{footnote}{0}
\setcounter{page}{0}
\newpage

The Standard Model (SM) of electroweak and strong interactions has been probed
at high-energy colliders for energies $\simlt$ 200 GeV, and confirmed with
an accuracy $\simlt$ 1~\%. In spite of this fact, we know that the SM cannot 
be a fundamental theory describing all interactions  because of its 
inability to give an answer to a 
number of more fundamental questions, as e.g. the hierarchy 
problem or how to consistently include gravity at the quantum level. 
For that reason extensions of the SM have been proposed and are nowadays
the object of experimental searches. 
In particular the minimal supersymmetric extension of the SM (MSSM), 
aiming to (technically) solve the hierarchy problem, is expected to arise as
an effective theory of some underlying fundamental supersymmetric theory 
valid at scales close to the Planck scale, as e.g. string of M-theory. 
However the knowledge of the fundamental theory is not enough to make the
link with the low energy SM, since it is known that supersymmetry is not
an exact symmetry of the nature and the mechanism of supersymmetry breaking
is for the moment unclear.

The presence of extra (compact) space dimensions is a common feature of any
fundamental theory valid at high scales. If their radii are of the order
of the Planck length, the corresponding excitations are superheavy and decouple
from the low-energy physics. However, if there is one or more large radii
they might have a number of phenomenological and theoretical 
implications~\cite{large1}-\cite{AQ1}. 
In particular there is recently an increasing interest
in understanding the role of large extra dimensions to describe the strong
coupling limit of string theory~\cite{HW}, to transmit
supersymmetry breaking between different  four-dimensional 
boundaries~\cite{H,Peskin}, to provide possible alternative
solutions to the hierarchy problem~\cite{hierarchy1,hierarchy2}, 
to modify the celebrated LEP
unification of gauge couplings~\cite{unificacion}, and even to 
explain possible modifications of gravitational measurements in the 
millimiter range~\cite{milimetro}.
Also large extra dimensions have been proved useful, and here is where we
focus our main interest, to provide a consistent and calculable 
supersymmetry breaking~\cite{AQ2,Emilian}.

In this letter we will follow the previous line of investigation
and explore the role of extra dimension(s) to provide the
source of new physics beyond the TeV scale and, at the
same time, to make the connection with the SM at present
energies. In particular we will present a simple $N=1$
supersymmetric model in five dimensions (5D) whose massless
sector, upon compactification of the fifth dimension on $S^1/{\bf Z}_2$,
is the MSSM while its massive modes rearrange in $N=2$ $D=4$
supermultiplets.
Furthermore, when the remaining $N=1$ $D=4$ supersymmetry 
of the massless modes is
broken by a Scherk-Schwarz~\cite{SS} mechanism on the fifth dimension, the
zero modes can be reduced to the non-supersymmetric Standard
Model, while the $N=2$ structure of the massive modes is in general
spoiled.

{\bf 1.} Here we introduce the model, based on a 5D theory
compactified on $S^1/{\bf Z}_2$. The model has $N=1$ supersymmetry. Non-chiral
matter, as the gauge and Higgs sector of the theory, lives in (5D) the
bulk of the fifth dimension, while chiral matter, the three generations
of quarks and leptons and their superpartners, live on the 4D boundary
(the fixed points of the orbifold $S^1/{\bf Z}_2$). This structure can arise
from string theories~\cite{large2} and has been recently considered for 
theoretical and phenomenological 
considerations~\cite{Peskin,hierarchy1,unificacion}.

In five dimensions, the vector supermultiplet 
$(V_M,\lambda^i_L,\Sigma)$
of an $SU(N)$ gauge theory consists
of a vector boson $V_M$ ($M=0,\dots,3,5$), 
a real scalar $\Sigma$ and two bispinors
$\lambda^i_L$ ($i=1,2$), all in the adjoint representation of $SU(N)$.
The 5D lagrangian is given by \footnote{The five dimensional
$N=1$ lagrangian can be deduced 
from the $N=2$ four dimensional lagrangian of Ref.~\cite{Sohnius}. We define
$V_M=gV^\alpha_M T^\alpha$, $\Sigma=g\Sigma^\alpha T^\alpha$ and 
$\lambda=g\lambda^\alpha T^\alpha$ where $T^\alpha$ are the generators 
of  $SU(N)$ with Tr$[T^\alpha T^\beta]=\delta^{\alpha\beta}/2$.}
\begin{equation}
{\cal L}=\frac{1}{g^2}{\rm Tr}\Big[-\frac{1}{2}{F_{MN}^2+|D_{M}\Sigma|^2+
i\overline\lambda_i\gamma^M D_M\lambda^i-
\overline\lambda_i}[\Sigma,\lambda^i]\Big]\, ,
\label{lagra1}
\end{equation}
where $\lambda^i$ is a symplectic-Majorana spinor
\begin{equation}                                                                                                                                
\lambda^i=\left(
\begin{array}{c}
\lambda^i_L \\
\epsilon^{ij}\;\overline\lambda_{Lj} 
\end{array}
\right)\, .
\label{symplectic}
\end{equation}
The 5D matter supermultiplet, ($H_i,\Psi$), consists 
of two scalar fields, $H_i$ ($i=1,2$), and a Dirac fermion 
$\Psi=(\Psi_L,\Psi_R)^T$. 
We will consider two matter supermultiplets, ($H^a_i$, $\Psi^a$) ($a=1,2$),
that we will associate with the two Higgs doublets of the MSSM.
The $N=1$ lagrangian for the matter supermultiplets interacting with the
vector supermultiplet is given by
\begin{eqnarray}
{\cal L}&=&|D_{M}H^a_i|^2+
i \overline\Psi_a\gamma^M D_M\Psi^a-
(i\sqrt{2}H^{\dagger i}_a \overline \lambda_i \Psi^a+h.c.)-
\overline\Psi_a\Sigma\Psi^a\nonumber\\
&-&
 H^{\dagger i}_a\Sigma^2H^{a}_i-
\frac{g^2}{2}\sum_{m,\alpha}[H^{\dagger i}_a (\sigma^{m})^{j}_{i}T^\alpha
H^{a}_j]^2\, ,
\label{lagra2}
\end{eqnarray}
where  $\sigma^m$ are the Pauli matrices.
The lagrangians of Eqs.~(\ref{lagra1}) and (\ref{lagra2}) have an 
$SU(2)_R\times SU(2)_H$ global symmetry.  
Under the $SU(2)_R\times SU(2)_H$ symmetry the fermionic fields transform
as doublets, $\lambda^i\in$({\bf 2},{\bf 1}), $\Psi^a\in$({\bf 1},{\bf 2}), 
while Higgs
bosons transform as bidoublets $H^a_i\in$({\bf 2},{\bf 2}). 
The rest of the fields in the vector multiplet are singlets.

Since all interactions in the lagrangians (\ref{lagra1}) and (\ref{lagra2})
are gauge interactions the model based on the gauge group
$SU(3)\times SU(2)\times U(1)$ is easily constructed from 
these expressions. It
contains 5D vector multiplets in the adjoint representation of
$SU(3)\times SU(2)\times U(1)$ 
[({\bf 8},{\bf 1},0)+({\bf 1},{\bf 3},0)+({\bf 1},{\bf 1},0)]
and two Higgs hypermultiplets in the representation
[({\bf 1},{\bf 2},1/2)+({\bf 1},{\bf 2},--1/2)]. 
The chiral matter is located on the
boundary and contains the usual chiral $N=1$ 4D multiplets. 

We will reduce the theory from five to four dimensions  by compactifying on
$S^1/{\bf Z}_2$, a circle with the identification $y\rightarrow -y$.
The transformation of the fields under the discrete parity ${\bf Z}_2$  
is determined by the interactions of Eqs.~(\ref{lagra1}) and (\ref{lagra2}).
We find
\begin{equation}
\Phi(x^\mu,y)\rightarrow \eta\Phi(x^\mu,-y)\,  ,
\label{paridad}
\end{equation}
where the states with $\eta=\pm 1$ are given in Table 1.
\begin{center}
\begin{tabular}{|c|c|c|c|c|c|} \hline
\multicolumn{3}{|c|}{$\eta=+1$} &\multicolumn{3}{|c|}{ $\eta=-1$}\\ \hline
$V_\mu$ & $H^2_2 $ & $H^1_1$ & $V_5,\Sigma$ & $H^2_1$ & $H^1_2$ \\
$\lambda^1_L$  & $\Psi^2_L$ & $\Psi^1_R$ & $\lambda^2_L$ & $\Psi^2_R$ &
$\Psi^1_L$ \\ \hline
\end{tabular}   
\vspace{0.5cm}

Table 1
\end{center}
We will call even the fields with $\eta=+1$ and odd those with
$\eta=-1$. Notice that we have rearranged fields in Table 1 in 
components of $N=1$ $D=4$ supersymmetric multiplets that are disposed
along the same column.

Compactifying on $S^1/{\bf Z}_2$ 
is equivalent to compactifying on a circle and imposing the discrete 
parity ${\bf Z}_2$.
In the $S^1$ compactification the fields can be Fourier expanded as 
\begin{equation}
\Phi(x^\mu,y)=\sum^\infty_{n=-\infty}e^{iny/R}\Phi^{(n)}(x^\mu)\,  ,
\label{kknp}
\end{equation}
where $R$ is the $S^1$ radius.
After integration with respect to the fifth dimension, the 
four dimensional theory consists  of a  tower of KK excitation
$\Phi^{(n)}(x^\mu)$  with mass $n/R$.
By imposing the ${\bf Z}_2$ parity,  the Fourier expansion is now given by
\begin{eqnarray}
\Phi_+(x^\mu,y)&=&\sum^\infty_{n=0}\cos \frac{ny}{R}\, 
\Phi^{(n)}(x^\mu), \ \ \ 
{\rm for\ the\ even\ fields}\, ,\nonumber\\
\Phi_-(x^\mu,y)&=&\sum^\infty_{n=1}\sin \frac{ny}{R}\, 
\Phi^{(n)}(x^\mu), \ \ \  
{\rm for\ the\ odd\ fields}\, .
\label{expansion}
\end{eqnarray}
Then the ${\bf Z}_2$ symmetry projects out   
half of the tower of KK modes of Eq.~(\ref{kknp}).
Each level of the KK-excitations form massive $N=2$ $D=4$ supermultiplets.
The $V_5^{(n)}$ field is eaten by the vector $V^{(n)}_\mu$ to become massive
while $\lambda^{1\,(n)}_L$ and $\lambda^{2\,(n)}_L$ 
become the components of a massive Dirac fermion.
These fields together with $\Sigma^{(n)}$ form an $N=2$ vector multiplet.
$H^{a\,(n)}_i$ and $\Psi^{a\,(n)}$ form two $N=2$ hypermultiplets.
The structure of $N=2$ supermultiplets is displayed in Table 2.
\begin{center}
\begin{tabular}{|c|c|c|} \hline
Vector multiplets &\multicolumn{2}{|c|}{ Hypermultiplets}\\ \hline
$V_\mu^{(n)}$, $\Sigma^{(n)}$ & $H_1^{1\,(n)}$, $H_2^{1\,(n)}$ &
$H_1^{2\,(n)}$, $H_2^{2\,(n)}$ \\
$\lambda_L^{1\,(n)}$, $\lambda_L^{2\,(n)}$ & $\Psi^{1\,(n)}$ & 
$\Psi^{2\,(n)}$ \\  \hline
\end{tabular}   
\vspace{0.5cm}

Table 2
\end{center}
At the massless level ($n=0$), however,  only the even fields, 
see Table 1, are left in the theory and we  have an $N=1$
supersymmetric theory.
Therefore the massless spectrum of this
model has just the MSSM content.

Notice that, by the process of compactification over $S^1/{\bf Z}_2$, the net
number of towers with states $\Phi^{(n)}$ ($n=-\infty,\dots,\+\infty$) is
divided by two. In fact the two towers defined by the 5D fields
$\Phi_+$ and $\Phi_-$ ($n\geq 0$) give rise, after compactification, to the
single tower $\Phi$ ($n\in {\bf Z}$) defined by
$\Phi=\Phi_+-i\Phi_-$,  i.e.
\begin{equation}
\Phi^{(\pm n)}=\frac{1}{2}\left\{\Phi_+^{(n)}\pm \Phi_-^{(n)}\right\}
,\ (n\geq 0).
\label{modos}
\end{equation}

{\bf 2.} In order to get, at the massless level, only
the SM fields we have to break supersymmetry.
We will use the Scherk-Schwarz (SS) mechanism~\cite{SS} that consists in
imposing  to the (superpartner)  5D fields 
a nontrivial periodic condition under a 2$\pi$ translation of the 
fifth dimension:
\begin{equation}
\Phi(x^\mu,y+2\pi)=e^{2\pi iqT}\Phi(x^\mu,y)\,  ,
\label{ss}
\end{equation}
where $T$  must be  a generator of a global symmetry of the five dimensional
theory \cite{SS} under which the field $\Phi$ transforms with charge $q$.
The requirement (\ref{ss})  implies a $y$-dependence 
for the fields different from that in eq.~(\ref{expansion}):
\begin{equation}
\Phi(x^\mu,y)=e^{iqTy/R}\widetilde \Phi(x^\mu,y)\,  ,
\label{ss2}
\end{equation}
where $\widetilde \Phi(x^\mu,y)$ are periodic  functions  
$\widetilde \Phi(x^\mu,y+2\pi)=\widetilde \Phi(x^\mu,y)$ and have the 
same Fourier expansion
as in Eq.~(\ref{expansion}).
Eq.~(\ref{ss2}) implies a nontrivial 
$y$-dependence for the $n=0$ mode that leads to 
a mass term in the four dimensional theory.

We will consider that $e^{2\pi iqT}$  belongs to 
$SU(2)_R\times SU(2)_H$. 
Since the theory is compactified on $S^1/{\bf Z}_2$, we must require 
\begin{equation}
{\bf Z}_2 e^{iqTy/R}=e^{iqT(-y)/R}{\bf Z}_2\,  .
\label{condition}
\end{equation}
This condition guarantees that $e^{iqTy/R}\widetilde\Phi$ has the same 
${\bf Z}_2$-transformation as $\widetilde\Phi$.   
The transformation ${\bf Z}_2$ on the 
$SU(2)_R$ and $SU(2)_H$ doublets, $\lambda^i$ and $\Psi^a$,
is respectively given by
\begin{equation}
{\bf Z}_2=\mp\pmatrix{
1 & 0 \cr
0 & -1 \cr
}\otimes \gamma_5\ ,
\label{z2}
\end{equation}
where the matrix $\mp\sigma_3$ in (\ref{z2}) acts on
$SU(2)_R$ or $SU(2)_H$ indices, while $\gamma_5$ acts on the
spinor indices.
Therefore the
condition (\ref{condition}) 
implies $T=\sin\theta\sigma_1+\cos\theta\sigma_2$~\cite{AQ2}
where $\sigma_i$ are the Pauli matrices representing the $SU(2)_{R,H}$ 
generators.
For simplicity we will only consider $\theta=0$.
Eq.~(\ref{ss2}) will imply
\begin{eqnarray}
\left (\matrix{
\lambda^1\cr
\lambda^2\cr
}\right )
&=&e^{iq_R\sigma_2y/R}
\left (\matrix{
\widetilde\lambda^1\cr
\widetilde\lambda^2\cr
}\right )\, ,
\nonumber\\ 
\left (\matrix{
\Psi^1\cr
\Psi^2\cr
}\right )
&=&e^{iq_H\sigma_2y/R}
\left (\matrix{
\widetilde\Psi^1\cr
\widetilde\Psi^2\cr
}\right )\, ,
\nonumber\\ & & \nonumber \\
\pmatrix{
H^1_1 & H^1_2 \cr
H^2_1 & H^2_2 \cr
}
&=&e^{iq_H\sigma_2y/R} 
\pmatrix{
\widetilde H^1_1 & \widetilde H^1_2 \cr
\widetilde H^2_1 & \widetilde H^2_2 \cr
}
e^{-iq_R\sigma_2y/R} \,  .
\label{ss3}
\end{eqnarray}

Using  the $y$-dependence of the fields given in
Eq.~(\ref{ss3}), we obtain, after integrating with respect 
to the fifth dimension $y$,
the following mass spectrum for $n\not= 0$:
\begin{eqnarray}
{\cal L}_{m}&=&
\frac{1}{R}\left\{\left (
\lambda^{1\,(n)}_L\, 
\lambda^{2\,(n)}_L
\right )
\pmatrix{
q_R & n \cr
n & q_R\cr
}\left (\matrix{
\lambda^{1\,(n)}_L\cr \lambda^{2\,(n)}_L
\cr
}\right )\right.\nonumber \\ & & \nonumber \\
&+ & \left.
\left (
\overline\Psi^{1\,(n)}_L\,
\overline\Psi^{2\,(n)}_L
\right )
\pmatrix{n&
-q_H  \cr
q_H&-n\cr
}\left (\matrix{
\Psi^{1\,(n)}_R\cr \Psi^{2\,(n)}_R\cr
}\right )+h.c.\right\}\\ & & \nonumber \\
-&{\displaystyle \frac{1}{R^2}} &\left (
H^{(n)\dagger}_0\, 
  H^{(n)\dagger}_2\, 
H^{(n)\dagger}_1\,  
H^{(n)\dagger}_3 \right )
\left (\matrix{
n^2+q_-^2&  -2inq_-&  &  \cr
2inq_- & n^2+q_-^2 &  &  \cr
 &  & n^2+q_+^2 & -2nq_+ \cr
 &  & -2nq_+ & n^2+q_+^2 \cr
}\right )\left (\matrix{
H^{(n)}_0\cr
H^{(n)}_2\cr
H^{(n)}_1\cr
H^{(n)}_3
}\right )\, ,\nonumber 
\end{eqnarray}
where we have 
redefined the scalar fields as  $H^a_i=H_\mu(\sigma^\mu)^a_i$,
$\sigma^\mu\equiv(1,\vec{\sigma})$, and~\footnote{Notice that with this
definition $H_0$ and $H_3$ are even, while $H_1$ and $H_2$ are odd
fields.} $q_{\pm}=q_H\pm q_R$.
Therefore the $n$-KK modes are now given by two Majorana fermions
$(\lambda^{1\,(n)}_L\pm \lambda^{2\,(n)}_L)$
with masses $|n\pm q_R|$, two  Dirac fermions,  $(\Psi^{1\, (n)}
\pm\Psi^{2\, (n)})$
with masses $|n\pm q_H|$, and four scalars, $(H^{(n)}_0\pm i\;H^{(n)}_2)$ and 
$(H^{(n)}_1\pm H^{(n)}_3)$ with masses $|n\pm(q_R-q_H)|$
and $|n\pm(q_R+q_H)|$ respectively. Of course, the mass spectrum of the fields
$V^{(n)}_\mu$, $V_5^{(n)}$ and $\Sigma^{(n)}$ 
is not modified by the SS compactification,
since they are singlets under $SU(2)_R\times SU(2)_H$.

For $n=0$, we have
\begin{eqnarray}
{\cal L}_{m}&=&\frac{1}{R}\left\{q_R\lambda^{1\;(0)}_L\lambda^{1\;(0)}_L+
q_H\overline\Psi^{2\;(0)}_L\Psi^{1\;(0)}_R+h.c.\right\}\nonumber\\
&-&\frac{1}{R^2}\left\{(q_R-q_H)^2\left|H_0^{(0)}\right|^2+
(q_R+q_H)^2\left|H_3^{(0)}\right|^2\right\}\, .
\end{eqnarray}
The massless spectrum now consists of only the $n=0$ mode of the vector fields 
$V^\mu$. Nevertheless  a massless scalar Higgs can be obtained if either of
the following conditions are satisfied:
\begin{itemize}
\item   
$q_R-q_H=n$, 
\item
$q_R+q_H=n$.
\end{itemize}

\noindent with $n=0,\pm 1,\pm 2,...$.
Therefore the massless spectrum of the model after 
the SS compactification can be reduced to the SM
with one or two Higgs doublets.
For example, for $q_R-q_H=0$ and $q_R+q_H\not =0$
we have a single massless scalar $H^{(0)}_0$.
We would like to associate this scalar with the SM Higgs.
Nevertheless we find that its self-interacting 
quartic coupling, given by the D-term potential 
\begin{equation}
V_D=g^2\sum_\alpha\varepsilon^{mnp}\left[H_0T^\alpha H_m^\dagger +i\; H_n
T^\alpha H_p^\dagger
+h.c. \right]^2  \ ,
\label{Dterms}
\end{equation}
is zero. This is because $H_0=H^1_1+H^2_2$ is a flat 
direction of the $D$-term contribution, as it is obvious from
Eq.~(\ref{Dterms}). 
Therefore $H_0$ has, at the tree level, a flat potential. Of course,
quantum effects will lift the flat direction since 
supersymmetry is broken, and we can expect that
they will induce a nontrivial minimum for the scalar Higgs.
Notice that the quantum structure 
of these theories is quite different from ordinary softly broken 
supersymmetric theories.
In our theory, the full tower of massive KK must be included at the one-loop
level. We leave this for further investigation.
A second alternative is to consider $q_R-q_H=0$ and $q_R-q_H=1$.
In this case one has two massless scalars $H^{(0)}_0$ and 
$H^{(1)}_1-H^{(1)}_3$. 
This scenario corresponds to a two Higgs doublet sector
similar to that of the MSSM.

Finally, the family sector of this theory resides in the four dimensional
boundary and then do not have KK excitations. Since the theory is $N=1$ 
supersymmetric, there are massless scalars associated to each fermion 
(the squarks and sleptons). Nevertheless, supersymmetry is broken in the bulk
and those scalar will get masses at the one-loop level. 
The massless spectrum will only correspond to the fermions of the SM.

\end{document}